\documentclass{article}
\usepackage{graphicx}
\usepackage{verbatim}
\usepackage{subfigure}
\usepackage{caption}
 \usepackage{wrapfig}

\usepackage{algo}  
\usepackage{url}  
\usepackage[utf8]{inputenc}  
\usepackage{subfigure}  
\usepackage{latexsym}  
\usepackage{amsmath} 
\usepackage{amssymb} 
\usepackage{algorithm}
\usepackage[left]{lineno}

\newtheorem{lemma}{Lemma} 

\newtheorem{theorem}{Theorem}

\newtheorem{definition}{Definition}

\renewcommand{\epsilon}{\varepsilon}
\renewcommand{\phi}{\varphi}

\begin{document} 
\bibliographystyle{plain}
 \title{\vspace*{-2cm}Indexing and querying color sets of images}

\author{
 Djamal Belazzougui\thanks{Department of Computer Science, FI-00014 University of Helsinki, Finland}
 \and
 Roman Kolpakov\thanks{Lomonosov Moscow State University, Moscow, Russia}
 \and
 Mathieu Raffinot\thanks{LIAFA, Universit\'e Paris Diderot--Paris 7, France}}
 \maketitle






\begin{abstract}
We aim to study the set of color sets of continuous regions
of an image given as a matrix of $m$ rows over
$n\geq m$ columns where each element in the matrix is an integer from
$[1,\sigma]$ named a {\em color}. 

The set of distinct colors in a region is called fingerprint.  
We aim to compute, index and query the fingerprints of all rectangular regions named rectangles. 
The set of all such fingerprints is denoted by ${\cal F}$.
A rectangle is {\em maximal} if it
is not contained in a greater rectangle with the same 
fingerprint. The set of all locations of maximal rectangles is denoted
by $\mathcal{L}.$ We first explain how to determine all the
$|\mathcal{L}|$ maximal locations with their fingerprints in expected
time $O(nm^2\sigma)$ using a Monte Carlo algorithm (with polynomially
small probability of error) or within deterministic
$O(nm^2\sigma\log(\frac{|\mathcal{L}|}{nm^2}+2))$ time. We then show
how to build a data structure which occupies $O(nm\log
n+\mathcal{|L|})$ space such that a query which asks for all the
maximal locations with a given fingerprint $f$ can be answered in time
$O(|f|+\log\log n+k)$, where $k$ is the number of maximal locations
with fingerprint $f$.
If the query asks only for the presence of the fingerprint, then the
space usage becomes $O(nm\log n+|{\cal F}|)$ while the query time
becomes $O(|f|+\log\log n)$. We eventually consider the special case
of squared regions (squares).
\end{abstract}



\section{Introduction}


In this paper, we are interested in studying and indexing the set of
all sets of distinct colors of continuous regions of a given image in
order to quickly answer to several queries on this set, for instance,
does there exist at least one region with a given set of colors and if
so, what are the positions of all such regions in the image?

\begin{figure}
\centering
$
\begin{array}{ccccc}
a_{m\;1}& a_{m\;2}&\ldots & a_{m\;n-1}& a_{m\;n}\\
a_{m-1\;1}& a_{m-1\;2}&\ldots & a_{m-1\;n-1}& a_{m-1\;n}\\
\vdots &\vdots &\ddots&\vdots &\vdots\\
a_{2\;1}& a_{2\;2}&\ldots & a_{2\;n-1}& a_{2\;n}\\
a_{1\;1}& a_{1\;2}&\ldots & a_{1\;n-1}& a_{1\;n}\\
\end{array}
$
\caption{An image}
\label{image}
\end{figure}

Also the considered indexing structures and algorithms are a good
approach towards efficient image comparisons and clustering based
on color sets. To our knowledge, this is the first time such a
problem is formalized and analyzed, and to build a general well
founded algorithmic framework we begin by formalizing our notions.

We consider an image as a matrix $M$ of $m$ rows over $n$ columns
(see Fig. \ref{image}) where each $a_{i,j}$ is an integer from
$[1,\sigma]$~\footnote{ We assume that $\sigma\leq nm$. If this is
  not the case, then we can build a dictionary data structure that
  occupies $O(nm)$ space and that can remap in constant time distinct
  characters from the original alphabet to distinct integers in
  $[1,nm]$.}. This matrix is called an image and henceforth, we will
refer to the integers stored in the matrix by colors or characters.

To design our indexing structures and algorithms, we extend the
definition of fingerprints (or sets of distinct characters), initially
defined on sequences \cite{AALS03,CYHW07,CYHW11,KR08b}, to images.
While classical pattern matching approaches on images have already
been studied ~\cite{AF92,AmirLS03,Ba78}, this paper is the first (to
our knowledge) to focus on the character sets. We assume below that
$m\le n$. We denote by $\langle i_0,i_1; j_0, j_1\rangle $ where
$i_0\le i_1$ and $j_0\le j_1$, the rectangle in~$M$ bounded by the
$i_0$-th and $i_1$-th rows and $j_0$-th and $j_1$-th columns including
these rows and columns. We also denote by $f \langle i_0, i_1; j_0,
j_1 \rangle$ the set of distinct colors contained in the rectangle
$\langle i_0, i_1; j_0, j_1 \rangle$. This set is called the {\em
  fingerprint} of $\langle i_0, i_1; j_0, j_1 \rangle$.

\begin{definition}
\label{defmax}
A rectangle in an image is {\em maximal} if it is not contained in a
greater rectangle with the same fingerprint.
\end{definition}

\begin{figure}[htb]
  \centering
\includegraphics[width=8cm]{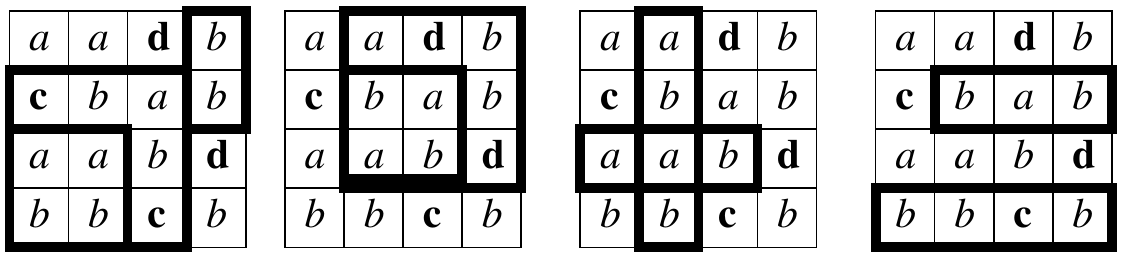}
\caption{Example of maximal rectangles (bold frontiers) in a small image}
 \label{example}
\end{figure}

In other words a rectangle is maximal if any extension of the
rectangle in one of the four directions will add at least one color
not present in the rectangle. Figure \ref{example} shows an example of
an image containing many maximal rectangles.

In this article, as a first step toward our general goal of indexing
all color sets of continuous regions, we focus on indexing and
answering queries on fingerprints of all maximal rectangles of the
input image. Given a fingerprint $f$, a maximal rectangle with
the fingerprint $f$ is called a {\em maximal location} of $f$. We
denote by $\mathcal{L}$ the set of all maximal locations, and by
$\mathcal{F}$ the set of all distinct fingerprints of rectangles in
the image. It is easy to see that $|{\cal F}|\le |{\cal L}|\le nm^2\sigma$.
All our results assume the standard RAM model with word
size $w=\Omega(\log (n+\sigma))$ and with all standard arithmetic and
logic operations (including multiplication) taking constant time. We
prove below the following theorems:

\begin{theorem}
Given an image of $m$ rows by $n\geq m$ columns, we can determine all the $|\mathcal{L}|$
maximal locations with their fingerprints in expected
time $O(nm^2\sigma)$ with a Monte Carlo algorithm (with polynomially
small probability of error) or within deterministic
$O(nm^2\sigma\log(\frac{|\mathcal{L}|}{nm^2}+2))$ time.
\end{theorem}

Note that the total deterministic time is $O(nm^2\sigma\log\sigma)$ in
the worst case (when we have $\mathcal{|L|}=\Theta(nm^2\sigma)$), but is only
$O(nm^2\sigma)$ (which is as good as the Monte Carlo time)
as long as $\mathcal{|L|}\leq O(nm^2)$.

\begin{theorem}
\label{thm:DS}
Given an image of $m$ rows by $n \geq m$ columns, 
whose set  ${\cal L}$ of all maximal locations and set ${\cal F}$  
of associated fingerprints have been already determined, we
can build a data structure which occupies space $O(nm\log n+\mathcal{|L|})$ such
that a query which asks for all the maximal locations
with a given fingerprint $f$ can be answered in time $O(|f|+\log\log
n+k)$, where $k$ is the number of maximal locations with fingerprint $f$.
If the query asks only for the presence of the fingerprint,
then the space usage becomes $O(nm\log n+|{\cal F}|)$ while the
query time becomes $O(|f|+\log\log n)$.
\end{theorem}

The construction times of the data structures mentioned in Theorem~\ref{thm:DS}
are respectively $O(nm\log n\log\log n+\mathcal{|L|})$ and 
$O(nm\log n\log\log n+\mathcal{|F|})$.

We eventually consider the special case in which only 
squared regions called squares are considered
instead of rectangles. 
We develop a specialized faster algorithm for this case.\\


\subsection{Notations and tools}


 Let $\langle i_0, i_1; j_0, j_1
\rangle$ be a
rectangle. For this rectangle, the $i_0$-th row is called the bottom
row, the $i_1$-th row is called the top row, the $j_0$-th column is
called the left column, and the $j_1$-th column is called the right
column. We also define the following notions derived from Definition
\ref{defmax}: a rectangle $\langle i_0, i_1; j_0, j_1
\rangle$ is {\em maximal to
  the left} (resp. {\em to the right}) if rectangle $\langle i_0, i_1; j_0-1,
j_1\rangle$ (resp. $\langle i_0, i_1; j_0, j_1+1\rangle $) doesn't have the same
fingerprint, and is {\em maximal to the bottom} (resp. {\em to the
  top}) if rectangle $\langle i_0-1, i_1; j_0, j_1\rangle $ (resp. $\langle i_0, i_1+1;
j_0, j_1\rangle$) doesn't have the same fingerprint. It is obvious that a
rectangle is maximal if and only if it is maximal in all the
directions. One of our solutions uses the following lemma by
Muthukrishnan~\cite{Mu02}:

\begin{lemma}
\label{lemma:color_1D_DS}

Given a sequence of colors $T[1,n]$ each chosen from the same alphabet
of colors $[1,\sigma]$, in time $O(n)$ we can preprocess the sequence into
a data structure which occupies $O(n)$ space so that given any
range $[i,j]$ we can find the set of all distinct colors
occurring in $T[i..j]$ in time $O(k)$ where $k$ is the number 
of reported colors. Moreover the data structure
reports the colors ordered by their last occurrence, returning 
the last occurrence of each color. 
\end{lemma}

\noindent
And also the following lemma from~\cite{LW13}:
\begin{lemma}
\label{lemma:color_2D_DS}
Given a set of $n$ colored points (with colors from $[1,\sigma]$) 
stored in a grid of $U$ columns by
$U$ rows, we can build a data structure which occupies $O(n\log n)$ space so that we can 
find the set of all distinct colors
occurring in any given rectangle in time $O(\log\log U+k)$, 
where $k$ is the number of distinct colors in the rectangle. 
A color may be reported up to $c$ times, for a known  
constant $c$. 
\end{lemma}
The lemmata above assume the same RAM model as ours with
$w=\Omega(\log (U+n+\sigma))$. It has been confirmed to 
us by one of the authors of~\cite{LW13} that the data structure 
of Lemma~\ref{lemma:color_2D_DS}  can be built in time $O(n\log n\log \log n)$. 

\section{Determination of all fingerprints and maximal locations}

We note that a (highly) naive algorithm which would try all possible
rectangles and explicitly compute the fingerprint of each rectangle
would take time $\Omega(n^2m^3)$~\footnote{We have $O(m^2n^2)$ rectangles
which we can process in a certain order so that we can determine the 
set of colors contained in every rectangle as the union of the $O(m)$ colors 
of the last column of the rectangle with the set of colors of 
another smaller rectangle that was previously processed.}. In the following, we show that
this cost can be reduced to just about $O(nm^2\sigma)$.  

Let, for some fixed $i_0$ and $i_1$, $R(i_0, i_1)$ be the set of all
rectangles with $i_0$-th bottom row and $i_1$-th upper row. Let, for
some fixed $j_0$ and $j_1$, $R(i_0, i_1; j_0, *)$ be the set of all
rectangles from $R(i_0, i_1)$ with $j_0$-th left column, and $R(i_0,
i_1; *, j_1)$ be the set of all rectangles from $R(i_0, i_1)$ with
$j_1$-th right column. For each $i_0$, $i_1$, and $j_0$ we construct
the sequence $\varphi (i_0, i_1; j_0)$ consisting of
distinct colors and end markers as follows. 

Let $r$ be the longest rectangle from $R(i_0, i_1; j_0, *)$ which is
maximal to the left and to the right (if it exists). Then the subsequence 
formed by all distinct colors of $\varphi (i_0, i_1; j_0)$ 
in left-to-right order is actually the sequence of all distinct colors from
the fingerprint of~$r$ satisfying the following property: if in this
sequence a color~$c$ is before a color~$b$ then the leftmost occurrence
of~$c$ in $\langle i_0, i_1; j_0, n\rangle$ is not to the right of the leftmost
occurrence of~$b$ in this rectangle. By definition, each maximal
rectangle from $R(i_0, i_1; j_0, *)$ is a subrectangle of~$r$, so
the set of colors in the fingerprint of such maximal rectangle is a
prefix of the sequence $\varphi (i_0, i_1; j_0)$. Then we
include in $\varphi (i_0, i_1; j_0)$ after the last color of this
prefix the end marker for this maximal rectangle. It is obvious that
distinct maximal rectangles from $R(i_0, i_1; j_0, *)$ have distinct
fingerprints, and thus distinct maximal rectangles correspond to
distinct end markers.

\subsection{Computing all $\varphi (i_0, i_1; j_0)$}
\label{allphi}


\subsubsection{Data structures}

For computing $\varphi (i_0, i_1; j_0)$ the following data structures
are used.  \\

\noindent
{\bf SLL}. 
The sequence of last
colors ($\mbox{SLL}$) for a string $p$ of colors is the sequence $(p_{i_1}, i_1), (p_{i_2},
i_2),\ldots , (p_{i_k}, i_k)$ where $p_{i_1}, p_{i_2},\ldots , p_{i_k}$ are all
colors contained in~$p$, and $i_j$ are the rightmost position of
color~$p_{i_j}$ in~$p$, and $i_1<i_2<\ldots<i_k$. We denote the $\mbox{SLL}$ 
for the string $a_{i,1}a_{i,2}\ldots a_{i,j}$ (the segment of the line 
number $i$ that spans columns $1$ to $j$) 
by $\mbox{SLL}\langle i; j\rangle$. \\

\noindent
For any rectangle~$r$ we also consider two following related
data structures. \\

\noindent
{\bf SLC}.
The sequence of last columns
($\mbox{SLC}$) for a rectangle $r$ is a sequence $(l_1, k_1),$ $(l_2, k_2),$ $\ldots
,$ $(l_s, k_s)$ with $l_1<l_2<\ldots <l_s$ and 
where $l_j$ is the number of a column containing at
least one rightmost occurrence of some color in $r$, 
and $k_j$ is the number of distinct colors
whose rightmost occurrences are contained in column $l_j$ (for
convenience we will assume that colors of the input alphabet which
have no occurrences in the rectangle are contained in 0-th column,
i.e. if there exist such colors then $\mbox{SLC}$ has first item $(0,
k)$ where $k$ in the number of such colors).\\

\noindent
{\bf LLP}.
The array of last color pointers ($\mbox{LLP}$) for a rectangle $r$, 
is an array of size $\sigma$ which for each color~$c$ contains the pointer 
to the item of $\mbox{SLC}$ corresponding to the column containing the rightmost occurrence(s) of~$c$
in~$r$. 

The $\mbox{SLC}$ sequence and $\mbox{LLP}$ array for $\langle i_0,
i_1; 1, j\rangle$ are denoted by $\mbox{SLC}\langle i_0, i_1; j\rangle$ and
$\mbox{LLP}\langle i_0, i_1; j\rangle$.
To support efficient updates, the sequences $\mbox{SLC}$ and $\mbox{SLL}$
are implemented using doubly linked lists.\\

\begin{figure}[htb]
\centering
$
\begin{array}{ccccccccccc}
& 1:& 2:& 3:& 4:& 5:& 6:& 7:& 8:& 9:& 10: \\
6: & b & g & d & i & f & f & e & e & c & g  \\
5: & e & d & e & i & f & h & e & e & a & i  \\
4: & i & f & d & b & b & i & i & i & e & e  \\
3: & a & e & j & d & b & j & i & a & h & e  \\
2: & e & f & h & a & i & e & b & b & e & f  \\
1: & c & a & b & e & g & f & i & b & g & i  \\
\end{array}
$
\caption{An instance of a rectangle image.}
\label{rectexample}
\end{figure}
\noindent
Figure \ref{rectexample} shows an instance of a rectangular image. In this example:
\begin{itemize}

\item $SLL\langle 1; 10\rangle = (c, 1), (a, 2), (e, 4), (f, 6), (b, 8), (g, 9), (i, 10)$

\item $SLL\langle 6; 10\rangle = (b, 1), (d, 3), (i, 4), (f, 6), (e, 8), (c, 9), (g, 10)$

\item $SLC\langle 2, 5; 10\rangle = (0, 2), (4,1), (6, 1), (8, 1), (9, 2), (10, 3)$

\item $LLP\langle 2, 5; 10\rangle$ contains for colors $a, b, c, d, e, f, g, h, i, j$
the pointers to $(9, 2)$, $(8, 1)$, $(0, 2)$, $(4,1)$, $(10, 3)$,
$(10, 3)$, $(0, 2)$, $(9, 2)$, $(10, 3)$, $(6, 1)$ respectively.
\end{itemize}

\subsubsection{Algorithm: main frame}
We present the following algorithm for computing all the sequences
$\varphi (i_0, i_1; j_0)$ for fixed $i_0$ and $i_1$. This algorithm
processes sequentially all columns of the rectangle
$\langle i_0, i_1; 1, n\rangle$ from first to $n$-th. 
Before processing the $j$-th column we assume
that we have computed the sequences $\mbox{SLL}\langle i_0-1; j-1\rangle$,
$\mbox{SLL}\langle i_1+1; j-1\rangle$, $\mbox{SLC}\langle i_0, i_1; j-1\rangle$ and the array
$\mbox{LLP}\langle i_0, i_1; j-1\rangle$. We compute also the fingerprint
$f\langle i_0, i_1; j, j\rangle$ for the $j$-th column of $\langle i_0, i_1; 1, n\rangle$. Let
$
\mbox{SLL}\langle i_0-1; j-1\rangle=\{(a'_1, i'_1), (a'_2, i'_2),\ldots , (a'_{k'}, i'_{k'})\},
$
$
\mbox{SLL}\langle i_1+1; j-1\rangle=\{(a''_1, i''_1), (a''_2, i''_2),\ldots , (a''_{k''}, i''_{k''})\},
$
$
\mbox{SLC}\langle i_0, i_1; j-1\rangle=\{(l_1, k_1), (l_2, k_2),\ldots , (l_s, k_s)\}
$ and $f\langle i_0, i_1; j, j\rangle=f$. The processing of the $j$-th column has two
stages.  
\subsubsection{First stage}
At the first stage we append to the sequences $\varphi (i_0,
i_1; j_0)$ the end markers for all maximal rectangles from $R(i_0, i_1; *,
j-1)$.
\paragraph{Left and right borders.}  
To that end, we initially find the
leftmost among the columns of $\langle i_0, i_1; 1, j-1\rangle$ 
that contain the rightmost occurrences 
of colors from~$f$ and the item of $\mbox{SLC}\langle i_0, i_1; j-1\rangle$ 
corresponding to this column
(this can be done in $O(\sigma )$ time, using $\mbox{LLP}\langle i_0, i_1;
j-1\rangle$). Let $l_t$ be the number of this column. Note that
$l_s=j-1$. It is easy to see that, if $t<s$, in the set $R(i_0, i_1;
*, j-1)$ the rectangles with left columns in positions $l_t+1,
l_{t+1}+1,\ldots , l_{s-1}+1$ are all the rectangles which are maximal
to the left and to the right (if $t=s$, there are no such rectangles).
\paragraph{Bottom and top borders}
 However, these rectangles may not be maximal since they may not be
 maximal to the bottom or to the top. Thus we need to check
 additionally for these rectangles the maximality to the bottom and
 to the top. For any color~$c$ denote by~$l(c)$ the number of the
 column in $\langle i_0, i_1; 1, j-1\rangle$ which contains the rightmost
 occurrences of the color~$c$ in $\langle i_0, i_1; 1, j-1\rangle$. Using
 $\mbox{SLC}\langle i_0, i_1; j-1\rangle$ and $\mbox{LLP}\langle i_0, i_1; j-1\rangle$, the
 value $l(c)$ can be computed in constant time. Note that a rectangle
 $\langle i_0, i_1; j_0, j-1\rangle$ from $R(i_0, i_1; *, j-1)$ is maximal to the
 bottom if and only if in $\mbox{SLL}\langle i_0-1; j-1\rangle$ there exists a
 color $a'_p$ such that $l(a'_p)<j_0\le i'_p$. Thus, if $l(a'_p)<
 i'_p$, the color $a'_p$ yields the interval $[l(a'_p)+1 .. i'_p]$ of
 numbers of left columns for rectangles from $R(i_0, i_1; *, j-1)$
 which are maximal to the bottom, and this interval can be computed in
 constant time. Therefore, in $O(\sigma )$ time we can construct the
 ordered sequence $I'$ of all the intervals yielded by the colors
 from $\mbox{SLL}\langle i_0-1; j-1\rangle$ (if some of these intervals are
 adjacent or overlapped we merge them in one interval in $I'$), and we
 have that $I'$ represents the set of all numbers of left columns for
 rectangles from $R(i_0, i_1; *, j-1)$ which are maximal to the
 bottom. Using $\mbox{SLL}\langle i_1+1; j-1\rangle$, $\mbox{SLC}\langle i_0, i_1; j-1\rangle$
 and $\mbox{LLP}\langle i_0, i_1; j-1\rangle$, we construct analogously the ordered
 sequence $I''$ of intervals presenting the set of all numbers of left
 columns for rectangles from $R(i_0, i_1; *, j-1)$ which are maximal
 to the top.\\ 

We then construct the ordered sequence $I$ of intervals which are
intersections of intervals from $I'$ and $I''$. Since both $I'$ and
$I''$ contain no more than $\sigma$ intervals, $I$ contains no more
than $2\sigma$ intervals, and therefore can be computed in $O(\sigma
)$ time. The sequence $I$ represents the set of all numbers of left
columns for rectangles from $R(i_0, i_1; *, j-1)$ which are maximal to
the bottom and to the top. Thus, to find all maximal rectangles from
$R(i_0, i_1; *, j-1)$, for each $q=t, t+1,\ldots , s-1$ we have to
check if the possible position $l_q+1$ for a left column of such
rectangle is contained in an interval of $I$ (this checking can be done
in total $O(\sigma )$ time). If it is the case, we insert in $\varphi
(i_0, i_1; l_q+1)$ the end marker for the maximal rectangle with
$(j-1)$-th right column. 

\subsubsection{Second stage}

At the second stage of the processing of the $j$-th column
we add to the sequences $\varphi (i_0, i_1; j_0)$ new colors from~$f$
which can be contained in maximal rectangles from $R(i_0, i_1)$ which
contain $j$-th column. To that end, we firstly compute for each
$l_q$-th column from $\mbox{SLC}\langle i_0, i_1; j-1\rangle$ the number $k'_q$ of
distinct colors from~$f$ whose rightmost occurrences are contained in
this column (it can be done in $O(\sigma )$ time, using
$\mbox{SLC}\langle i_0, i_1; j-1\rangle$ and $\mbox{LLP}\langle i_0, i_1; j-1\rangle$). 

We call the $l_q$-th column from $\mbox{SLC}\langle i_0, i_1; j-1\rangle$ {\it
  feasible} if $k'_q<k_q$. Using the numbers $k'_q$, we can compute in
$O(\sigma)$ time the subsequence of feasible columns from
$\mbox{SLC}\langle i_0, i_1; j-1\rangle$. Using $\mbox{SLC}\langle i_0, i_1; j-1\rangle$ and
$\mbox{LLP}\langle i_0, i_1; j-1\rangle$, we can also compute in $O(\sigma)$ time
for each color~$c$ from~$f$ the index $t(a)$ such that the rightmost
occurrences of~$c$ are contained in $l_{t(a)}$-th column of $\langle i_0,
i_1; 1, j-1\rangle$. After computing these indexes for each color~$c$
from~$f$ we add the color~$c$ to $\varphi (i_0, i_1; l_q+1)$ for each
feasible $l_q$-th column from $\mbox{SLC}\langle i_0, i_1; j-1\rangle$ such that
$t(a)\le q$.
\subsubsection{Complexity.}
 The time complexity of the processing of the $j$-th column is $O(\sigma +S)$ where $S$ is
 the number of inserted colors. Thus the total time complexity of
 this procedure for all~$j=1, 2,\ldots, n$ is $O(n\sigma +\hat S)$
 where $\hat S$ is the sum of values $S$ of all sequences
 $\varphi (i_0, i_1; j_0)$. Since any sequence $\varphi (i_0, i_1;
 j_0)$ contains no more than $\sigma$ colors, we have $\hat S\le
 n\sigma$, i.e. the total time complexity of this procedure for
 all~$j$ is $O(n\sigma )$.


\subsubsection{Algorithm: data structures update from column $j$ to $j+1$.}
After processing the $j$-th column we compute the structures
$\mbox{SLL}\langle i_0-1; j\rangle$, $\mbox{SLL}\langle i_1+1; j\rangle$, $\mbox{SLC}\langle i_0, i_1;
j\rangle$ and $\mbox{LLP}\langle i_0, i_1; j\rangle$, required for processing the $j+1$-th
column, from $\mbox{SLL}\langle i_0-1; j-1\rangle$, $\mbox{SLL}\langle i_1+1; j-1\rangle$,
$\mbox{SLC}\langle i_0, i_1; j-1\rangle$ and $\mbox{LLP}\langle i_0, i_1; j-1\rangle$. It can
be done obviously in $O(\sigma )$ time. 


\subsubsection{Last step}
Once we have processed all the columns, we include in the sequences $\varphi (i_0, i_1; j_0)$ the end markers for
maximal rectangles from $R(i_0, i_1; *, n)$, using the structures
$\mbox{SLL}\langle i_0-1; n\rangle$, $\mbox{SLL}\langle i_1+1; n\rangle$, $\mbox{SLC}\langle i_0, i_1;
n\rangle$ and $\mbox{LLP}\langle i_0, i_1; n\rangle$. This procedure is similar to the
procedure described above except that in this case we consider for
possible positions of left columns the position~1 and positions
$l_q+1$ for each $l_q$-th column from $\mbox{SLC}\langle i_0, i_1; n\rangle$.

\subsubsection{Algorithm's complexity.} 

The time complexity for 
computing all sequences $\varphi (i_0, i_1; j_0)$ 
is clearly $O(nm^2\sigma )$ excluding the time for the computation
of the sets~$f$ which is shown next.  

\subsection{Computing the all sets of colors: two variants}

We present below two variants for resolving the problem 
of effective computation of the sets~$f$. The first variant 
relies on Lemma~\ref{lemma:color_1D_DS}, the second is more complex 
but does not rely on any external data structure. Both variants run in $O(n m^2\sigma)$
time.\\


\noindent
{\bf First variant.} The first variant of resolving this problem uses the result stated
in Lemma~1 which allows to compute~$f$ in~$O(|f|)$ time, 
which is bounded by $O(\sigma)$. Thus, given 
$i_0$ and $i_1$, computing $\varphi (i_0, i_1; j_0)$ for 
all $j_0$ is bounded by $O(n\sigma)$ and repeating $O(m^2)$ 
such iterations for all pairs $i_0,i_1$ takes
$O(n m^2\sigma)$ time.\\

\noindent
{\bf Second variant.} The second variant avoids the direct computation of the set~$f$ 
by computing in parallel for each fixed~$i_0$ all the 
sequences $\varphi (i_0, i_1; j_0)$ for
any~$j_0$ and any $i_1\ge i_0$. In this case at the $j$-th step we
process at once the $j$-th columns sequentially in each rectangle $\langle i_0,
i_1; 1, n\rangle$ for $i_1=i_0, i_0+1,\ldots , m$. We assume that before
processing this step we have computed all sequences $\mbox{SLL}\langle i_0-1;
j-1\rangle, \mbox{SLL}\langle i_0; j-1\rangle, \mbox{SLL}\langle i_0+1; j-1\rangle,\ldots ,
\mbox{SLL}\langle m; j-1\rangle$. Note that the data structures $\mbox{SLC}\langle i_0,
i_1; j-1\rangle$ and $\mbox{LLP}\langle i_0, i_1; j-1\rangle$, required for processing the
$j$-th column in $\langle i_0, i_1; 1, n\rangle$, can be computed from the disposed
sequence $\mbox{SLL}\langle i_1; j-1\rangle$ and the structures $\mbox{SLC}\langle i_0,
i_1-1; j-1\rangle$ and $\mbox{LLP}\langle i_0, i_1-1; j-1\rangle$ used for processing the
$j$-th column in $\langle i_0, i_1-1; 1, n\rangle$ in $O(\sigma )$ time. Moreover,
the set $f\langle i_0, i_1; j, j\rangle$, required for processing the $j$-th column
in $\langle i_0, i_1; 1, n\rangle$, can be computed in constant time from the set
$f\langle i_0, i_1-1; j, j\rangle$, used for processing the $j$-th column in $\langle i_0,
i_1-1; 1, n\rangle$. Thus, after processing the $j$-th column in $\langle i_0, i_1-1;
1, n\rangle$ we can compute in $O(\sigma )$ time all the data structures,
required for processing the $j$-th column in $\langle i_0, i_1; 1, n\rangle$.
Moreover, after processing the $j$-th step each sequence $\mbox{SLL}\langle i;
j\rangle$, required for $(j+1)$-th step, can be computed in constant time from
the sequence $\mbox{SLL}\langle i; j-1\rangle$. So all sequences $\mbox{SLL}\langle i;
j\rangle$, required for $(j+1)$-th step, can be computed in $O(m)$ time. Thus,
the time required for computing all data structures at each step is
$O(m\sigma)$, i.e. the total time for computing all data structures in
this parallel procedure is $O(mn\sigma)$.

Since the whole algorithm requires $m$ such iterations, the whole
complexity is also $O(nm^2\sigma)$ time.

\subsection{Naming}
\label{sec:naming}

The naming technique is used to give a unique name to the fingerprint
(color set) of each maximal rectangle of an image. 
The technique which originated in~\cite{KMR72} was first adapted to the 
fingerprinting problem in~\cite{AALS03} and later its speed 
was improved in~\cite{DSST07}. 

\begin{figure}[htb]
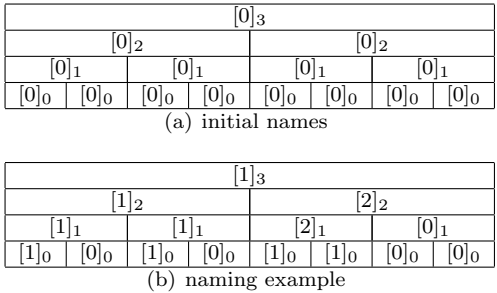

{\footnotesize
\begin{center}
 \captionsetup[subfigure]{aboveskip=+3pt,belowskip=+1pt}
\subfigure[initial names]{
$\begin{array}{|c|c|c|c|c|c|c|c|}
\hline
\multicolumn{8}{|c|}{[0]_3}\\
\hline
\multicolumn{4}{|c|}{[0]_2} & \multicolumn{4}{|c|}{[0]_2}\\
\hline
\multicolumn{2}{|c|}{[0]_1} & \multicolumn{2}{|c|}{[0]_1} &
\multicolumn{2}{|c|}{[0]_1} & \multicolumn{2}{|c|}{[0]_1}\\
\hline
[0]_0 & [0]_0 & [0]_0 & [0]_0 & [0]_0 & [0]_0 & [0]_0 & [0]_0\\
\hline
\end{array}$}
 \quad
\subfigure[naming example]{
$
\begin{array}{|c|c|c|c|c|c|c|c|}
\hline
\multicolumn{8}{|c|}{[1]_3}\\
\hline
\multicolumn{4}{|c|}{[1]_2} & \multicolumn{4}{|c|}{[2]_2}\\
\hline
\multicolumn{2}{|c|}{[1]_1} & \multicolumn{2}{|c|}{[1]_1} &
\multicolumn{2}{|c|}{[2]_1} & \multicolumn{2}{|c|}{[0]_1}\\
\hline
[1]_0 & [0]_0 & [1]_0 & [0]_0 & [1]_0 & [1]_0 & [0]_0 & [0]_0\\
\hline
\end{array}
$}
\end{center}
\caption{Naming initialization and instance.}
\label{namingexample}
}
\end{figure}

We assume for simplicity, but without loss of generality, that
$\sigma$ is a power of two. We consider a stack of $\log\sigma+1$
arrays $B_0\ldots B_{\log\sigma}$ on top of each other. The levels are numbered bottom-up
starting from 0. The lowest $B_0$, called the fingerprint table, contains
$\sigma$ names that are only $[0]_0$ or $[1]_0$, where a 
$[1]_0$ indicates the presence of the character. A name is indexed by
its level number. The array of any level $i>0$ contains
half the number of names of the array of level $i-1$ it is placed
upon. The highest array only contains a single name that will be the
name of the whole array. Such a name is called a fingerprint
name. Figure \ref{namingexample} shows a simple example with
$\sigma=8$, where the left shows the stack for an empty character
set and the right shows the stack for a non-empty one.

The names in the fingerprint table are only $[0]_0$ or $[1]_0$ and are
given. Each cell, $c$, of an upper array represents two disjoint 
and consecutive cells of the array it is placed upon, 
and thus a pair of two names. 
Conceptually, the naming is done in the following way: for each level 
going from the lowest to the
highest, if the cell represents a new pair of names, give this pair
a new name and assign it to the cell. If the pair has already been
named, place this name into the cell. In the example in
Figure \ref{namingexample}, the name $[1]_1$ is associated to $([1]_0,[0]_0)$
the first time this pair is encountered. The second time, this name is
directly retrieved.
The approach above can be directly implemented by
using a dictionary for level $i$ that stores all the pairs of names
from level $i$ associated with a name from level $i+1$. 
The dictionary is either kept using a binary search tree~\cite{AALS03} or 
a hash table~\cite{BKR13}. 
In~\cite{DSST07} the naming is only done at the end of the computation of the 
maximal locations. 
Here we show a slightly modified version of that naming 
adapted for our case. 

\paragraph{\bf Naming algorithm.} 
Before doing the naming, we first start by preprocessing the sequences 
$\varphi (i_0, i_1; j_0)$ by 
sorting (using radix sort) the subsequences of 
characters that lie between two end markers 
in the sequences (or before the first end marker). 
If $t$ is the number of characters to sort, then this phase
will take time $O(t(\frac{\log \sigma}{\log t}+1))$. 

Our naming will proceed in $\log\sigma$ phases. 
At phase $i\in[1..\log\sigma]$, we will determine all the cells 
at level $i$. Before the first phase, 
the stack of arrays $B_0\ldots B_\sigma$ is initialized 
at zero. 

At phase $1$, we build a unique list $L_1$ from
all the sequences $\varphi (i_0, i_1; j_0)$ processed in sequential order 
and using two global counters $C_0$ and $C_1$ (initially at $0$) as follows: 
in every list $\varphi (i_0, i_1; j_0)$, and 
for every subsequence of characters that lies between two 
consecutive end markers (or before the first one), 
we do the following: 
we set to one all the positions $B_0[\alpha]$ for every 
character $\alpha$ in the subsequence. We then scan the subsequence again 
and replace every character $\alpha$ by $\lceil\alpha/2\rceil$ 
(and potentially merge any two consecutive resulting characters 
if they are equal), generate a pair of pairs 
$((C_1,C_0),(B_0[2\alpha-1],B_0[2\alpha]))$ 
and increment $C_0$. After processing the subsequence we increment
the counter $C_1$ and reinitialize $C_0$ at $0$.

At the end of processing a sequence $\varphi (i_0, i_1; j_0)$, 
we reinitialize $B_0$ to zero by undoing all the operations 
done on $B_0$. 
After we have processed all the sequences $\varphi$,
we sort the elements in the generated pairs of pairs 
by their last components
(which are actually pairs of names from level $0$) and replace 
all equal last components with a unique sequential number
(a name for level $1$). We then again sort the resulting pairs, this 
time by their first components (the value of $(C_1,C_0)$) resulting in a list
$L_1$. which we partition into sublists according to the value 
of $C_1$. 
At this time, we have already built all the names at level $1$ 
and we can proceed to the phase number $2$ which builds all the names 
for level $2$. We first reset $C_1$ to $0$. 
We then scan all the sequences $\varphi (i_0, i_1; j_0)$
and the list $L_1$ in parallel and for a subsequence number $k$ 
(as before the subsequences are separated 
by end markers and their number is indicated by $C_1$) 
do the following two steps. 
In the first step, we apply the changes 
of $B_1$ indicated by the corresponding sublist in $L_1$. That is, 
if the character number $j$ in the subsequence is $\alpha$, 
we write in $B_1[\alpha]$ the last component 
(the name for level $1$) coming from the pair number $j$ in the sublist 
number $k$ of $L_1$. 

In the second step, we rescan again the subsequence $k$
and replace every character $\alpha$ by $\lceil\alpha/2\rceil$ 
(and removing duplicates), generate a pair $((C_1,C_0),(B_1[2\alpha-1],B_1[2\alpha]))$ 
and increment $C_0$ . After processing the subsequence $k$ we increment
the counter $C_1$ and reinitialize $C_0$ at $0$. 
At the end of processing a sequence $\varphi (i_0, i_1; j_0)$, 
we reinitialize $B_1$ to zero by undoing all the operations 
done on $B_1$. 
Everything else is exactly the same as in level number $1$. 
The sorting step will result in an array $L_2$ 
that allows to replace every unique pair of 
names from level $1$ by a name for level $2$. 

At any phase $i>2$, the algorithm is exactly the same as in level $2$, 
except that we now use $B_{i-1}$, $L_i$ and $L_{i-1}$ 
instead of $B_1$, $L_2$ and $L_1$. 

At the end of the phase number $\sigma$, every sublist of $L_\sigma$ 
will contain a single pair whose last component indicates the name of 
a maximal rectangle.

\paragraph{\bf Complexity.} 
We first bound the overhead due to the 
sorting phases. Sorting is first used on the $t$ characters 
in a subsequence and takes (recall that a subsequence stores all
the $t$ characters that are added to produce the fingerprint 
of a maximal rectangle)
$O(t(\frac{\log\sigma}{\log t}+1))\leq O(t(\log\frac{\sigma}{t}+1))$ time. 
It is also used twice for each phase $i$, in order 
to produce the list $L_i$. 


We can prove that the naming of each maximal location involving the addition of $t$ 
new colors is done in total time $O(t(\log\frac{\sigma}{t}+1))$. 
In this case all $|{\cal L}|$ maximal locations 
can be named in total time $O(m^2n\sigma \max\{1,\log (|{\cal L}|/nm^2)$
which will be faster than a naive naming when $|{\cal L}| = o(nm^2\sigma)$.

The trick is to view the stack as a full binary tree of 
height $\log(\sigma)$, where the $t$
leaves (the $t$ positions in the fingerprint table) 
are marked and the internal nodes in the
root-to-leaves paths leading to those $t$ leaves are also marked
(those nodes correspond to generated names).
Then the total number of marked nodes will be less than
$O(t(\log\frac{\sigma}{t}+1))$. This is obvious: consider the upper 
$\lceil\log t\rceil$ levels of the perfect binary tree.
The total number of nodes into these upper levels can not exceed 
$2^{\lceil\log t\rceil}\leq 2t$. The total number of internal nodes
in the lower $\log\sigma-\lceil\log t\rceil$ levels will
be clearly no more than 
$t(\log\sigma-\lceil\log t\rceil)=O(t\log\frac{\sigma}{t})$. 


Then, the naming will be done level by level starting from the bottom level
and taking in total $O(t(\log\frac{\sigma}{t}+1))$ time for a maximal location
whose fingerprint adds $t$ colors to the previous one. 
\label{pageproof}


We prove now a $O(m^2n\sigma\max\{1,\log\frac{|{\cal L}|}{nm^2}\})$ bound for the naming of all maximal locations.
For convenience, the $\log$ function will refer 
to the base $e$ logarithm instead of base $2$ (as is the case in 
the rest of the paper). 

We denote the number of color(s) added by the maximal
location number $i$ by $t_i$. We let $\ell=|\cal L|$ and $\sum_{i=1}^{\ell}t_i=T$. 
We have: 
$$\begin{array}{rl}
&\sum_{i=1}^{\ell} O\left(t_i(\log\left(\frac{\sigma}{t_i}\right)+1)\right)\\
&=\sum_{i=1}^{\ell} O\left(t_i\log\left(\frac{\sigma}{t_i}\right)\right)+\sum_{i=1}^{\ell}O\left(t_i\right)\\
&=\sum_{i=1}^{\ell} O\left(t_i\log\left(\frac{\sigma}{t_i}\right)\right)+O\left(T\right).
\end{array}
$$
Since $T=O(m^2n)$, in order to bound the naming time we only 
need to prove that $P=\sum_{i=1}^{\ell} O\left(t_i\log\left(\frac{\sigma}{t_i}\right)\right)\leq O(m^2n\sigma\max\{1,\log\frac{|{\cal L}|}{nm^2}\})$. 

We now apply Jensen's inequality to the convex function $f(x)=x\log x$. 
Recall that the Jensen inequality for a convex function $f(x)$ states that: 
$$
f\left(\frac{\sum_{i=1}^{n}x_i}{n}\right)\leq \frac{\sum_{i=1}^{n}f(x_i)}{n}
$$
By replacing $x_i$ by $t_i$, $f(x)$ by $x\log x$ and $n$ by $\ell$ we obtain:

$$\left(\frac{\sum_{i=1}^{\ell}t_i}{\ell}\right)\log\left(\frac{\sum_{i=1}^{\ell}t_i}{\ell}\right)\leq
\frac{\sum_{i=1}^{\ell}t_i\log t_i}{\ell}
$$
which simplifies to the following inequality:
$T\log \frac{T}{\ell}\leq \sum_{i=1}^{\ell}t_i\log t_i $. 

By replacing in the term $P$ we get:
$$
\begin{array}{rl}
&P=\sum_{i=1}^{\ell} O\left(t_i\log\left(\frac{\sigma}{t_i}\right)\right)\\
&=O\left(\left(\sum_{i=1}^{\ell}t_i\log\sigma\right)-\left(\sum_{i=1}^{\ell}t_i\log t_i\right)\right)\\
&\leq O\left(T\log\sigma-T\log\frac{T}{\ell}\right)\\
& =O\left(T\left(\log\frac{\sigma}{\frac{T}{\ell}})\right)\right)=O(\sigma g(T/\sigma))
\end{array}
$$

where $g(x)=x\log(\ell/x)$. Note that $g$
is  monotonically increasing for $x\le \ell/e$ and
monotonically decreasing for $x\ge \ell/e$ and is maximal 
at $x=\ell/e$. So we consider two cases:
\begin{enumerate}
\item $\ell/e\leq nm^2$, in this case  
$$\sigma g(T/\sigma)\leq \sigma g(\ell/e)=\sigma (\ell/e)\log e =O(\sigma \ell)=O(nm^2\sigma)$$

\item $\ell/e > nm^2$. Since $T\leq nm^2\sigma$ we have that $T/\sigma\leq nm^2<\ell/e$, and 
$$\sigma g(T/\sigma)\leq \sigma g(nm^2)=nm^2\sigma\log(\ell/nm^2)$$
\end{enumerate}
So in the both cases $P=\sigma g(T/\sigma)\leq O(nm^2\sigma\max\{1,\log\frac{|{\cal L}|}{nm^2}\})$
which concludes the analysis.

\subsection{Storing fingerprints}

We now show how to store the fingerprints efficiently in $O(|{\cal
  L}|+nm\log n)$ space such that a query for a fingerprint $f$ will take
only time $O(\log\log n+|f|+k)$, where $k$ is the number of maximal 
locations that correspond to the fingerprint.
 If a query asks just for the presence of a fingerprint,
without requiring to return its corresponding maximal locations, then
we can do with just query time $O(\log\log n+|f|)$ with space usage
reduced to just $O(|{\cal F}|+nm\log n)$.
We note that storing all the fingerprints could take much more space
than just $O(|{\cal F}|)$. This is unlike the one dimensional case in
which every fingerprint adds only one color to another
fingerprint and hence all the fingerprints can be coded with just
$O(1)$ space overhead per fingerprint (actually the overhead can 
even be reduced to just $2\log\sigma+O(1)$ bits per fingerprint)~\cite{BKR13}. 
In our case of two dimensional fingerprints, a fingerprint can add an 
arbitrary number of colors to another fingerprint. Hence coding a
fingerprint by just storing the differing colors with other
fingerprints would require too much space.
Our solution to use just $O(1)$ overhead per fingerprint is to index
the matrix using the data structure of Lemma~\ref{lemma:color_2D_DS}. This would
use space $O(nm\log (nm))=O(nm\log n)$. We also store in a hash
table all the hash values associated with all distinct fingerprints (perfect 
hash function). To each value, we associate a pointer to one of the rectangles 
having the corresponding fingerprint. More details on the hash function 
implementation are given in the next subsection. For now we assume that 
the perfect hash function can be evaluated in time $O(|f|)$ on a 
fingerprint $f$. Now, given a fingerprint $f$, a query will proceed in four 
steps: (1) First compute the hash value $h$ associated with $f$. This is
  done in time $O(|f|)$.
(2) Probe the hash table for the value $h$ and if found, retrieve
  the pointer to the rectangle associated with that hash value. Otherwise 
declare a failure and stop the query. 
(3) Retrieve all the colors occurring inside that
  rectangle but stop if there are more than $c|f|$ distinct colors (for some 
fixed and known constant $c$). If that was the case then the query is stopped with a failure. 
(4) Match the reported colors with the set $|f|$.

We give now more details on the four steps. The first step can
obviously be done in time $O(|f|)$. The second step is done in $O(1)$
time. Then third step can be done in time $O(|f|+\log\log (n+m))$ 
if the data structure of Lemma~\ref{lemma:color_2D_DS} is used (with
the constant $c$ as defined in the lemma). Eventually the last 
step can be done in $O(|f|)$ time as follows. We only use a bit-vector
of length $\sigma$ bits which will be reserved specifically for queries.
Initially, all the bits in the bit-vector are set to zero. For a given 
query consisting in a fingerprint $f$, we first set to one all the bits in the bit-vector
whose positions correspond to colors in the fingerprint. 
To check that the colors reported by the data structure
of Lemma~\ref{lemma:color_2D_DS} are the same as those of the fingerprint,
we start with a counter set to zero and then do the 
following for each retrieved color. 
First check if the corresponding bit in the bitvector
is set to one. If this is the case, we reset the bit to zero 
and increment the counter. 
After we have processed all the reported colors, we can report a success
if and only if the value of the counter is $|f|$. 
Eventually at the end of the query, and regardless of the result of
the query, we reset all the bits that were set to one during the query, 
in such a way that the bitvector is made only of zeros (by setting to zero
all the positions of the bits which were set to one during the query).
We have thus proved that existential queries can be answered in time 
$O(|f|+\log\log n)$ using space $O(|{\cal F}|+nm\log n)$. 
In order to support reporting queries, we will additionally store the list
of maximal locations (maximal rectangles) that correspond to each fingerprint. 
This adds $O(|{\cal L}|)$ space to the data structure. At query time, we only 
need to check that the colors in the first maximal location correspond to 
the given query fingerprint $f$ in time $O(|f|+\log\log n)$. 
Afterward, we can report the remaining maximal $k-1$ rectangles 
by traversing the list in time $O(k-1)$. 

\subsection{Hash table and probabilistic naming}
\label{sec:hashing}
In order to implement the hash table described in previous subsection, we 
can make use of a polynomial hash function (Rabin-Karp signatures~\cite{KR87})
so as to injectively reduce the fingerprints to integers of length $O(\log(|{\cal F}|))$. 
This approach was used and described in detail in~\cite{BKR13}. We only give a sketch
here. 
The used polynomial hash function $H$ is parametrized 
by a prime number $r$ from the range $[1..n^c]$ for some suitable constant $c$.
For a randomly chosen $r$ from the range, the hash functions
maps injectively the fingerprints
to integers in $[1,|{\cal F}|^{O(1)}]$ with high probability (probability 
$1-\frac{1}{|{\cal F}|^{\Theta(1)}}$). 
If it fails to do so, then we choose a different $r$ from the range 
until we find a hash function that maps all fingerprints to distinct integers. 
Then we can use any minimal perfect hashing scheme~\cite{FKS82} to map those 
integers to the final range $[1..|{\cal F}|]$. 
Note that evaluating a perfect hash function on any given fingerprint $f$ takes
$O(|f|)$ time as it consists in first computing the Rabin-Karp signature followed
by the computation of the perfect hash function applied on the obtained integer. 

The use of hashing provides an alternative naming that will work with high probability 
(Monte Carlo). Simply omit the phase described in section~\ref{sec:naming} and 
simply rely on the polynomial hash function $H$ to give distinct names 
to different fingerprints. The polynomial hash function will give 
distinct names to all fingerprints with high probability. The advantage of such an approach
is that the total complexity of the naming is $O(nm^2\sigma)$. This is because computing the 
hash value of a fingerprint that adds $t$ characters to another fingerprint given the hash value
of the latter can be done in time $O(t)$.

\section{Matching squares}

We can obtain a faster algorithm in case we are only aiming at
matching square maximal locations. A square is {\em maximal} if and
only if it is not included in a greater square having the same set of
colors. 



\subsection{Notations}

We  denote by $[i, j; k]$ the square $\langle i, i+k-1; j, j+k-1\rangle$ where
$a_{i, j}$ is the left bottom corner, $a_{i+k-1, j}$ is the left upper corner,
$a_{i, j+k-1}$ is the right bottom corner, and $a_{i+k-1, j+k-1}$ is the right upper corner.
By $f[i, j; k]$ we denote the fingerprint of $\langle i, i+k-1; j, j+k-1\rangle$ and we call the parameter~$k$
the {\it size} of this square. Let the $r$-th diagonal be the set of all points $a_{i, j}$ such that $i-j=r$.
A square is on a diagonal if its left bottom and right upper corners are in this diagonal.

We also denote by $S[i, j; *]$ the set of all squares with the left bottom corner $a_{i, j}$,
and by $S[*; i, j]$ the set of all squares with the right upper corner $a_{i, j}$. For any
two colors $a_{i', j'}$ and $a_{i'', j''}$ the distance between $a_{i', j'}$ and $a_{i'', j''}$ 
is $\min (|i'-i''|, |j'-j''|)$. By $\lceil i, j; k\rceil$ we denote the triangle with
corners $a_{i, j}$, $a_{i+k, j}$ and $a_{i+k, j+k}$, and by $\lfloor i, j; k\rfloor$ we denote
the triangle with corners $a_{i, j}$, $a_{i, j+k}$ and $a_{i+k, j+k}$. The $i$-th line of a triangle 
is the intersection of the $i$-th line of the input image with this triangle, and the $j$-th column of 
a triangle is the intersection of the $j$-th column of the input image with this triangle.

\subsection{Maximality conditions}

A square $[i, j; k]$ is maximal if it fulfills the 
square maximality conditions $L, $ $R,$ $ U,$ $ D,$ $ LU,$ $ LD,$ $ RU,$ $ RD$ 
defined as follows: 
\begin{itemize}
\item $L$ is true if $f\langle i, i+k-1; j-1, j+k-1\rangle\neq f[i, j; k]$
\item $R$ is true if $f\langle i, i+k-1; j, j+k\rangle\neq f[i, j; k]$
\item $U$ is true if $f\langle i, i+k; j, j+k-1\rangle\neq f[i, j; k]$
\item $D$ is true if $f\langle i-1, i+k-1; j, j+k-1\rangle\neq f[i, j; k]$
\item $LU$ is true if $a_{i+k, j-1}\notin f[i, j; k]$ and $f\langle i, i+k; j, j+k-1\rangle = f[i, j; k]$
\item $LD$ is true if $a_{i-1, j-1}\notin f[i, j; k]$
\item $RU$ is true if $a_{i+k, j+k}\notin f[i, j; k]$
\item $RD$ is true if $a_{i-1, j+k}\notin f[i, j; k]$ and if $f\langle i, i+k-1; j, j+k\rangle = f[i, j; k]$
\end{itemize}

Note that a square is maximal if and only if the condition
$$
(L\vee U\vee LU)\wedge (L\vee D\vee LD)\wedge
(R\vee U\vee RU)\wedge (R\vee D\vee RD)
$$
holds for the square.

For each $i$ and~$j$ we construct the following sequence $\hat\varphi (i; j)$ consisting
of distinct colors and end markers. Let $s_{i, j}$ be the greatest square in $S[i, j; *]$.
For any color~$c$ contained in $s_{i, j}$ the distance of~$c$ in $s_{i, j}$ is the minimal
distance between $a_{i, j}$ and occurrences of~$c$ in $s_{i, j}$. The color subsequence of
$\hat\varphi (i; j)$ is a sequence of all distinct colors contained in $s_{i, j}$ such that
if in this sequence a color~$c$ is before a color~$b$ then the distance of~$c$ in $s_{i, j}$
is not greater than the distance of~$b$ in $s_{i, j}$. Note that for any square from $S[i, j; *]$
the fingerprint of this square forms a starting segment of the considered color subsequence.
For any maximal square from $S[i, j; *]$ we insert in $\hat\varphi (i; j)$ after the last color 
of the corresponding segment the end marker for this maximal square (it is obvious that
distinct maximal squares have distinct end markers). Thus, to find all maximal squares on a fixed 
diagonal we have to compute all sequences $\hat\varphi (i; j)$ such that $a_{i, j}$ is in this diagonal.

\subsection{Computing  $\hat\varphi (i; j)$}

\subsubsection{Data structures}

For computing $\hat\varphi (i; j)$ the following data structures are used.\\

\noindent
{\bf SCT}. 
For the triangle 
$\lceil i, j; k\rceil$ we consider the two-way queue $\mbox{SCT}\lceil i, j; k\rceil$.
This is the sequence $(c_1, q_1), (c_2, q_2),\ldots , (c_s, q_s)$ such that 
$c_1<c_2<\ldots<c_s$ where $c_t$ is the number of a column of the triangle $\lceil i, j; k\rceil$
which contains occurrences of colors such that the square $[i+c_t-j, c_t+1; j+k+1-c_t]$ has no
occurrences of these colors. We will call such colors {\it respective} to $c_t$-th column
of the triangle $\lceil i, j; k\rceil$. $q_t$ is the number of distinct colors respective
to $c_t$-th column of the triangle $\lceil i, j; k\rceil$.\\ 

\noindent
{\bf PCT}.
We will also consider the array 
$\mbox{PCT}\lceil i, j; k\rceil$ which for each color~$c$ contains the pointer to the item
of $\mbox{SCT}\lceil i, j; k\rceil$ for the column of $\lceil i, j; k\rceil$ respective to~$c$
if such column exists; otherwise this pointer is undefined. \\ 

\noindent
{\bf SLT}.
For the triangle $\lfloor i, j; k\rfloor$
we consider the two-way queue $\mbox{SLT}\lfloor i, j; k\rfloor$. This is the sequence 
$(l_1, p_1), (l_2, p_2),\ldots , (l_s, p_s)$ such that $l_1<l_2<\ldots<l_s$ where $l_t$ 
is the number of a line of the triangle $\lfloor i, j; k\rfloor$ which contains occurrences 
of colors such that the square $[l_t+1, j+l_t-i; i+k+1-l_t]$ has no occurrences of these colors.
We will also call such colors {\it respective} to $l_t$-th line of the triangle $\lfloor i, j; k\rfloor$.\\ 

\noindent
{\bf PLT}.
Similarly to the array $\mbox{PCT}\lceil i, j; k\rceil$ for $\mbox{SCT}\lceil i, j; k\rceil$,
we consider also the array $\mbox{PLT}\lfloor i, j; k\rfloor$ for $\mbox{SLT}\lfloor i, j; k\rfloor$.\\ 

\noindent
{\bf SLD}.
Moreover, for each $i$ and~$j$ we consider the two-way queue $\mbox{SLD}\langle i, j\rangle$.
This is the sequence $(a_{i_1,i_1+r}, i_1), (a_{i_2,i_2+r}, i_2),\ldots ,(a_{i_s,i_s+r}, i_s)$
where $r=j-i$, $i_1<i_2<i_3<\ldots <i_s=i$, and $a_{i_t, i_t+r}$ has no occurrences in the
square with the left bottom corner $a_{i_t+1, i_t+r+1}$ and the right upper corner $a_{i, j}$.\\ 

\noindent
{\bf PLD}.
For $\mbox{SLD}\langle i, j\rangle$ we also consider array $\mbox{PLD}\langle i, j\rangle$ which for each color~$c$ contains 
the pointer to the item of $\mbox{SLD}\langle i, j\rangle$ with this color if such item exists; otherwise this 
pointer is undefined. \\ 

\noindent
{\bf SLL} and {\bf SLL}$^T$.
Besides the sequences $\mbox{SLL}\langle i; j\rangle$ we also use the sequences
$\mbox{SLL}^T\langle i; j\rangle$ which are the $\mbox{SLL}$ sequences for the string $a_{1,j}a_{2,j}\ldots a_{i,j}$.\\ 

\begin{figure}[htb]
\centering
$
\begin{array}{ccccccccc}
& 1:& 2:& 3:& 4:& 5:& 6:& 7:& 8: \\
8: & a & f & d & a & f & f & i & c  \\
7: & h & f & d & g & i & j & a & i  \\
6: & j & d & i & b & g & g & a & c  \\
5: & i & f & i & i & a & h & i & f  \\
4: & f & j & d & b & b & g & j & h  \\
3: & h & d & a & i & f & h & c & b  \\
2: & a & g & i & g & i & a & h & b  \\
1: & h & f & e & e & b & d & c & b  \\
\end{array}
$
\caption{An instance of a square.}
\label{squareexample}
\end{figure}

\noindent
Figure \ref{squareexample} shows an instance of a square. In this example:
\begin{itemize}
\item $SCT\lceil 2, 1; 6\rceil = (3, 1), (4, 1), (6, 2), (7, 1)$
\item $PCT\lceil 2, 1; 6\rceil$ contains for letters $b, d, f, i, j$
the pointers to $(4, 1)$, $(3, 1)$, $(6, 2)$, $(7, 1)$, $(6, 2)$ respectively
\item $SLT\lfloor 1, 2; 6\rfloor = (1, 1), (4, 1), (5, 1), (7, 1)$
\item $PLT\lfloor 1, 2; 6\rfloor$ contains for letters $b, e, h, i$
the pointers to $(4, 1)$, $(1, 1)$, $(5, 1)$, $(7, 1)$ respectively
\item $SLD\langle 8, 8\rangle = (b, 4), (g, 6), (a, 7), (c, 8)$
\item $PLD\langle 8, 8\rangle$ contains for letters $a, b, c, g$
the pointers to $(a, 7)$, $(b, 4)$, $(c, 8)$, $(g, 6)$ respectively
\end{itemize}

\subsubsection{Algorithm}

Consider the $r$-th diagonal of the input image where $r\ge 0$ (the case $r<0$ is processed in the same way).
To find all maximal squares on this diagonal, we need to compute all sequences 
$\hat\varphi (1; r+1), \hat\varphi (2; r+2),\ldots  $. For computing these sequences we process sequentially 
all colors $a_{i, i+r}$ where $i=1, 2,\ldots $. Before processing a color $a_{i, j}$, where $i=j+r$,
on this diagonal, we assume that we have computed the following structures: $\mbox{SCT}\lceil 2, 1+r; i-2\rceil$,
$\mbox{PCT}\lceil 2, r+1; i-2\rceil$, $\mbox{SLT}\lfloor 1, r+2; i-2\rfloor$, $\mbox{PLT}\lfloor 1, r+2; i-2\rfloor$,
$\mbox{SLD}\langle i-1, j-1\rangle$, $\mbox{PLD}\langle i-1, j-1\rangle$, $\mbox{SLL}\langle i; j-1\rangle$, and $\mbox{SLL}^T\langle i-1; j\rangle$. The processing 
of $a_{i, j}$ is done in two stages.

\paragraph{First stage.} At the first stage we insert in the computed sequences $\hat\varphi (l; r+l)$
the end markers for all maximal squares from $S[*; i-1, j-1]$. Note that each square $[i-k, j-k; k]$ 
from $S[*; i-1, j-1]$ is uniquely defined by its size~$k$. Thus, for each maximality condition 
$P\in\{L, R, D, U, LD, RU, LU, RD\}$ we can compute the characteristic set $\chi_P$ of sizes
of all squares from $S[*; i-1, j-1]$ which satisfy the condition~$P$. First of all, for any color~$c$,
using the structures  $\mbox{SCT}\lceil 2, 1+r; i-2\rceil$, $\mbox{PCT}\lceil 2, r+1; i-2\rceil$, 
$\mbox{SLT}\lfloor 1, r+2; i-2\rfloor$, $\mbox{PLT}\lfloor 1, r+2; i-2\rfloor$, $\mbox{SLD}\langle i-1, j-1\rangle$, 
and $\mbox{PLD}\langle i-1, j-1\rangle$, we can compute in constant time the maximal size $K_c$ of a square from $S[*; i-1, j-1]$
which has no occurrences of color~$c$. Thus we can compute in constant time the set $\chi_{RU}=[1 .. K_{a_{i, j}}]$.
Further, if 
$$
\mbox{SLD}\langle i-1, j-1\rangle=(a_{i_1,i_1+r}, i_1), (a_{i_2,i_2+r}, i_2),\ldots ,(a_{i_s,i_s+r}, i_s)
$$
then we have obviously (taking into account that $i_s=i-1$) 
$$
\chi_{LD}=\{i-i_{s-1}-1, i-i_{s-2}-1,\ldots, i-i_1-1\}
$$
Now let 
$$
\mbox{SCT}\lceil 2, 1+r; i-2\rceil = (c_1, q_1), (c_2, q_2),\ldots , (c_s, q_s)
$$
Then obviously
$$
\chi_{L}=\{j-c_s-1, j-c_{s-1}-1,\ldots , j-c_1-1\}
$$
In the symmetrical way we can compute $\chi_{D}$ from $\mbox{SLT}\lfloor 1, r+2; i-2\rfloor$.
To compute $\chi_{U}$, note that a square $[i-k, j-k; k]$ from $S[*; i-1, j-1]$ satisfies
the condition~$U$ if and only if in $\mbox{SLL}\langle i; j-1\rangle$ there exists an item $(c_t, j_t)$
such that $j-j_t\le k\le K_{c_t}$. Therefore, each item $(c_t, j_t)$ of $\mbox{SLL}\langle i; j-1\rangle$
such that $j-j_t\le K_{c_t}$ yields the interval $[j-j_t .. K_{c_t}]$ to $\chi_{U}$, and
$\chi_{U}$ has no other intervals. Moreover, each of these intervals $[j-j_t .. K_{c_t}]$
can be computed in constant time. Thus, $\chi_{U}$ consists of no more than $\sigma$
intervals and can be computed in $O(\sigma )$ time. In the symmetrical way, $\chi_{R}$
also consists of no more than $\sigma$ intervals and can be computed in $O(\sigma )$ time
from $\mbox{SLL}^T\langle i-1; j\rangle$. Computing $\chi_{LU}$ is based on the following considerations.
Let $[i-k, j-k; k]$ be a square from $S[*; i-1, j-1]$ satisfying the condition $LU$.
So $a_{i, j-k-1}\notin f[i-k, j-k; k]=f\langle i-k, i; j-k, j-1\rangle$, i.e. the color $a_{i, j-k-1}$
is not contained in the string $a_{i, j-k}a_{i, j-k+1}\ldots a_{i, j-1}$. Thus $a_{i, j-k-1}$
is the rightmost occurrence of this color in the string $a_{i, 1}a_{i, 2}\ldots a_{i, j-1}$,
i.e. $a_{i, j-k-1}$ is contained in $\mbox{SLL}\langle i; j-1\rangle$. Hence, to compute $\chi_{LU}$,
we can check only values $k=j-j_t-1$ for each item $(c_t, j_t)$ from $\mbox{SLL}\langle i; j-1\rangle$.
More precisely, for each item $(c_t, j_t)$ from $\mbox{SLL}\langle i; j-1\rangle$ we check if the color 
$c_t$ has occurrences in $[i-k, j-k; k]$ where $k=j-j_t-1$ (since $K_{c_t}$ can be computed
in constant time, it can be checked in constant time). If it is the case, we include~$k$ in $\chi_{LU}$.
Thus, $\chi_{LU}$ contains no more than $\sigma$ values and can be computed in $O(\sigma )$ time.
In the symmetrical way $\chi_{RD}$ contains no more than $\sigma$ values and can be computed 
in $O(\sigma )$ time from $\mbox{SLL}^T\langle i-1; j\rangle$. Thus all characteristic sets for the
considered maximality conditions $L, R, D, U, LD, RU, LU, RD$ contain no more than $\sigma$
values or intervals and can be computed in $O(\sigma )$ time. Note that each value~$k$ can be
considered as interval~$[k .. k]$, so without loss of generality we can assume that all 
characteristic sets for the maximality conditions consist of no more than $\sigma$ intervals.
Thus the characteristic set
$$
\chi_{M}=(\chi_{L}\cup\chi_{U}\cup\chi_{LU})\cap (\chi_{L}\cup\chi_{D}\cup\chi_{LD})\cap
(\chi_{R}\cup\chi_{U}\cup\chi_{RU})\cap (\chi_{R}\cup\chi_{D}\cup\chi_{RD})
$$
of sizes $k$ of all maximal squares $[i-k, j-k; k]$ from $S[*; i-1, j-1]$ consists of $O(\sigma)$ intervals
and can be computed in $O(\sigma )$ time. After computing $\chi_{M}$, for each~$k$ from $\chi_{M}$ 
we insert in $\hat\varphi (i-k; j-k)$ the end marker for the maximal square $[i-k, j-k; k]$.
Thus, the time complexity of the first stage computations is $O(\sigma +S)$ where $S$ is the
number of maximal squares in $S[*; i-1, j-1]$. So the overall time complexity of the second stage 
computations is $O(nm\sigma +\hat S)$ where $\hat S$ is the total number of maximal squares.
Since $\hat S\le nm\sigma$, we have that the overall time complexity of the second stage 
computations is $O(nm\sigma )$.

\paragraph{Second stage.} At the second stage of processing $a_{i, j}$ we add to the computed sequences $\hat\varphi (i-k; j-k)$ 
new colors which are not contained in squares $[i-k, j-k; k]$ but are contained in squares $[i-k, j-k; k+1]$.
We make the following computations. Note that for each color~$c$, using $\mbox{SLL}\langle i; j-1\rangle$, $\mbox{SLL}^T\langle i-1; j\rangle$,
$a_{i, j}$ and $K_{c}$, we can compute in constant time the maximal size $\hat K_c$ of a square from $S[*; i, j]$
which has no occurrences of~$c$. Then, if $\hat K_c\le K_c$, we add the color~$c$ to each of the sequences
$\hat\varphi (i-K_{c}; j-K_{c})$, $\hat\varphi (i-K_{c}+1; j-K_{c}+1)$,\ldots , $\hat\varphi (i-\hat K_{c}; j-\hat K_{c})$.
The time complexity of these computations is $O(\sigma +S)$ where $S$ is the number af added colors.
Thus the overall time complexity of the second stage computations is $O(nm\sigma +\hat S)$ where $\hat S$
is the total number of colors in all sequences $\hat\varphi (i; j)$. So, since $\hat S\le nm\sigma$,
the overall time complexity of the second stage computations is $O(nm\sigma )$.

Note also that the data structures $\mbox{SCT}\lceil 2, 1+r; i-2\rceil$, $\mbox{PCT}\lceil 2, r+1; i-2\rceil$, 
$\mbox{SLT}\lfloor 1, r+2; i-2\rfloor$, $\mbox{PLT}\lfloor 1, r+2; i-2\rfloor$, $\mbox{SLD}\langle i-1, j-1\rangle$, 
$\mbox{PLD}\langle i-1, j-1\rangle$, required for processing the color $a_{i, j}$ can be computed in $O(\sigma )$ time
from the data structures $\mbox{SCT}\lceil 2, 1+r; i-3\rceil$, $\mbox{PCT}\lceil 2, r+1; i-3\rceil$, 
$\mbox{SLT}\lfloor 1, r+2; i-3\rfloor$, $\mbox{PLT}\lfloor 1, r+2; i-3\rfloor$, $\mbox{SLD}\langle i-2, j-2\rangle$, 
$\mbox{PLD}\langle i-2, j-2\rangle$, $\mbox{SLL}\langle i-1; j-2\rangle$, and $\mbox{SLL}^T\langle i-2; j-1\rangle$, used for processing the
previous color $a_{i-1, j-1}$. Thus the overall time complexity of computing the structures
$\mbox{SCT}\lceil 2, 1+r; i-2\rceil$, $\mbox{PCT}\lceil 2, r+1; i-2\rceil$, $\mbox{SLT}\lfloor 1, r+2; i-2\rfloor$, 
$\mbox{PLT}\lfloor 1, r+2; i-2\rfloor$, $\mbox{SLD}\langle i-1, j-1\rangle$, and $\mbox{PLD}\langle i-1, j-1\rangle$
is $O(nm\sigma )$. The "bottleneck" of this algorithm is the computation of the remaining
structures $\mbox{SLL}\langle i; j-1\rangle$ and $\mbox{SLL}^T\langle i-1; j\rangle$. To resolve this problem, we can
use the result stated in Lemma \ref{lemma:color_1D_DS} which allows to compute these structures in $O(\sigma )$ time.
Another way of resolving this problem is to process all diagonals in parallel. More precisely,
the algorithm makes $n$ steps for $j=1, 2,\ldots , n$, and at $j$-th step the colors
$a_{1, j}, a_{2, j},\ldots , a_{m, j}$ are sequentially processed. We assume that before $j$-th step, 
all data structures required for processing these colors, except the structures
$\mbox{SLL}^T\langle 1; j\rangle$, $\mbox{SLL}^T\langle 2; j\rangle$,\ldots , $\mbox{SLL}^T\langle m; j\rangle$ are already computed. 
In this case the sequence $\mbox{SLL}^T\langle i-1; j\rangle$, required for processing $a_{i, j}$,
can be computed in constant time from the sequence $\mbox{SLL}^T\langle i-2; j\rangle$, used for
processing the previous color $a_{i-1, j}$. Thus the overall time complexity of computing
all sequences $\mbox{SLL}^T\langle i-1; j\rangle$ is $O(mn)$. Moreover, each sequence $\mbox{SLL}\langle i; j-1\rangle$,
required at $j$-th step, can be computed in constant time from the sequence $\mbox{SLL}\langle i; j-2\rangle$,
used at $(j-1)$-th step. So the overall time complexity of computing all sequences $\mbox{SLL}\langle i; j-1\rangle$
is also $O(mn)$. Thus the total time complexity of the algorithm for finding all maximal squares
is $O(nm\sigma )$.\\

\subsection{Complexity}

\noindent
By combining the previous algorithm with the naming scheme of section
\ref{sec:naming}, we obtain the following theorem.

\begin{theorem}
Given an image of $m$ rows by $n\geq m$ columns, we can compute all the $|\mathcal{S}|$
maximal squares with 
the set ${\cal F}$ of all distinct fingerprints of these squares
in deterministic time $O(mn\sigma\log(\frac{|\mathcal{S}|}{nm}+2))$ or in Monte Carlo
algorithm in time $O(mn\sigma)$. We can build a data
structure which occupies space $O(nm\log n+|\mathcal{S}|)$ such that a query
which asks for all the maximal squares with a given
fingerprint $f$ can be answered in time $O(|f|+\log\log n+k)$,
where $k$ is the number of maximal squares. If the
query asks only for the 
presence of squares with a given fingerprint $f$, then the space
usage becomes $O(nm\log n+|{\cal F}|)$ while the query time becomes
$O(|f|+\log\log n)$.
\end{theorem}
The construction times of the data structures mentioned in the theorem
are respectively $O(nm\log n\log\log n+\mathcal{|S|})$ and 
$O(nm\log n\log\log n+\mathcal{|F|})$. 

\section*{Acknowledgements}
The authors wish to thank Kasper Green Larsen and Yakov Nekrich 
for confirming that the 2-dimensional range color reporting 
data structure of Lemma~\ref{lemma:color_2D_DS} can be built in time 
$O(n\log n\log \log n)$.

\small
\bibliography{fingerprint0}
\normalsize

\end{document}